\documentclass%
[aps,prb,twocolumn,footinbib,
showpacs,
superscriptaddress,
]{revtex4-1}
\usepackage{hyperref}
\usepackage{amsmath}
\usepackage{amsfonts}
\usepackage{amssymb}
\usepackage{graphicx}
\newcommand{\fig}[2]{\includegraphics[width=#1]{#2}}

\newcommand*{\rmd}{\mathrm{d}}

\newcommand*{\bv}[1]{\boldsymbol{#1}}

\begin{document}

\title{High-Q Nanomechanics via Destructive Interference of
  Elastic Waves}

\author{I.~Wilson-Rae} \email[]{ignacio.wilson-rae@ph.tum.de}
\affiliation{Technische Universit\"{a}t M\"{u}nchen, 85748
  Garching, Germany.}
\author{R.A.~Barton}
%\affiliation{Center for Materials Research, Cornell University,
%  Ithaca, NY, 14853 USA} 
\author{S.S.~Verbridge} 
%\affiliation{Center for Materials Research, Cornell University,
%  Ithaca, NY, 14853 USA} 
\author{D.R.~Southworth}
%\affiliation{Center for Materials Research, Cornell University,
%  Ithaca, NY, 14853 USA} 
\author{B.~Ilic}
%\affiliation{Center for Materials Research, Cornell University,
%  Ithaca, NY, 14853 USA} 
\author{H.G.~Craighead}
%\affiliation{Center for Materials Research, Cornell University,
%  Ithaca, NY, 14853 USA} 
\author{J.M.~Parpia}
\affiliation{Center for Materials Research, Cornell University,
  Ithaca, NY, 14853 USA} 

\date{\today}

\begin{abstract}
  Mechanical dissipation poses an ubiquitous challenge to the
  performance of nanomechanical devices. Here we analyze the
  support-induced dissipation of high-stress nanomechanical
  resonators. We develop a model for this loss mechanism and test it
  on Si$_{3}$N$_{4}$ membranes with circular and square
  geometries. The measured $Q$-values of different harmonics present a
  non-monotonic behavior which is successfully explained. For
  azimuthal harmonics of the circular geometry we predict that
  destructive interference of the radiated waves leads to an
  exponential suppression of the clamping loss in the harmonic
  index. Our model can also be applied to graphene drums under high
  tension.
\end{abstract}


\maketitle

Nanomechanical resonators offer great potential for practical device
applications that exploit their ultra-low mass and high frequencies
\cite{Craighead00Ekinci05}. Examples range from scaling scanning-probe
force microscopy and mass-sensing down to the molecular scale to
providing alternatives for radio frequency devices. These applications
share the desirability of high mechanical quality-factor ($Q$) that,
by virtue of narrow bandwidth, amounts to a better defined frequency
thus enhancing performance. In turn, measurements of mechanical
displacements with an imprecision below the standard quantum limit and
the preparation of ultracold motional states have already been
implemented with electromechanical
\cite{Teufel09,Rocheleau10,O'Connell10} and optomechanical systems
\cite{Schliesser09,Groblacher09}. These breakthroughs foreshadow the
possibility of realizing a ``quantum optics'' analogue involving a
macroscopic mechanical degree of freedom which would set a new stage
for fundamental tests and potential quantum devices
\cite{Armour02,Blencowe04,O'Connell10}. Once more mechanical
dissipation, as determined by the $Q$-value, plays a critical role in
such endeavors.

Though the mechanical $Q$-value may in general be influenced by
various mechanisms, in a small suspended structure that is
sufficiently clean and cold, internal losses induced by two-level
fluctuators
\cite{Mohanty02,Seoanez08Remus09,Southworth09,Venkatesan10} and
radiation of elastic waves into the substrate are likely to play
leading roles
\cite{Cross01Bindel05,Wilson-Rae08,Anetsberger08,Eichenfield09}. In
this respect, the reduction of the design-limited ``clamping loss''
induced by the coupling to the substrate will allow the use of
nanomechanical devices to probe the internal losses and quantify the
fundamental contributions of the constituent materials.  Furthermore,
with the advent of the use of stressed silicon nitride membranes, the
high $Q$-values of these devices have demonstrated that it is indeed
possible to attain low mechanical losses in nanoresonators
\cite{Southworth09,Verbridge06,Verbridge07,Verbridge08,Thompson08,Wilson09,Unterreithmeier10}.

In this letter, we present and test a model that captures the energy
loss that occurs due to elastic wave radiation at the periphery of
these high-stress resonators. We show that this mechanism is
significant in state of the art structures and is strongly influenced
by interference effects. We compare the results of our model to
measurements of the resonant modes of two configurations, a single
``drum resonator'' and a composite array of drum resonators that
effectively realizes a square membrane [cf.~Fig.~1 (a) and (b)]. We
examine the harmonics of these structures and accurately account for
much of the variation in the corresponding $Q$-values. Our analysis
reveals that certain types of modes are inherently resilient to
clamping loss as a result of destructive interference of the radiated
waves. Thus, we provide insight into resonators that might be realized
and yield better $Q$-values in the future. On general grounds, the
fact that the relevant stress at the resonator-support contact scales
at least linearly with frequency combined with the 3D nature of the
support, lead to the naive expectation that the dissipation ($1/Q$)
due to elastic-wave radiation should increase as one considers higher
harmonics [cf.~Eq.~(\ref{Q})]. In dramatic contrast, we find that for
the harmonics of a circular membrane the clamping loss is
exponentially suppressed as the number of radial nodal lines
increases.

\begin{figure*}
\fig{\textwidth}{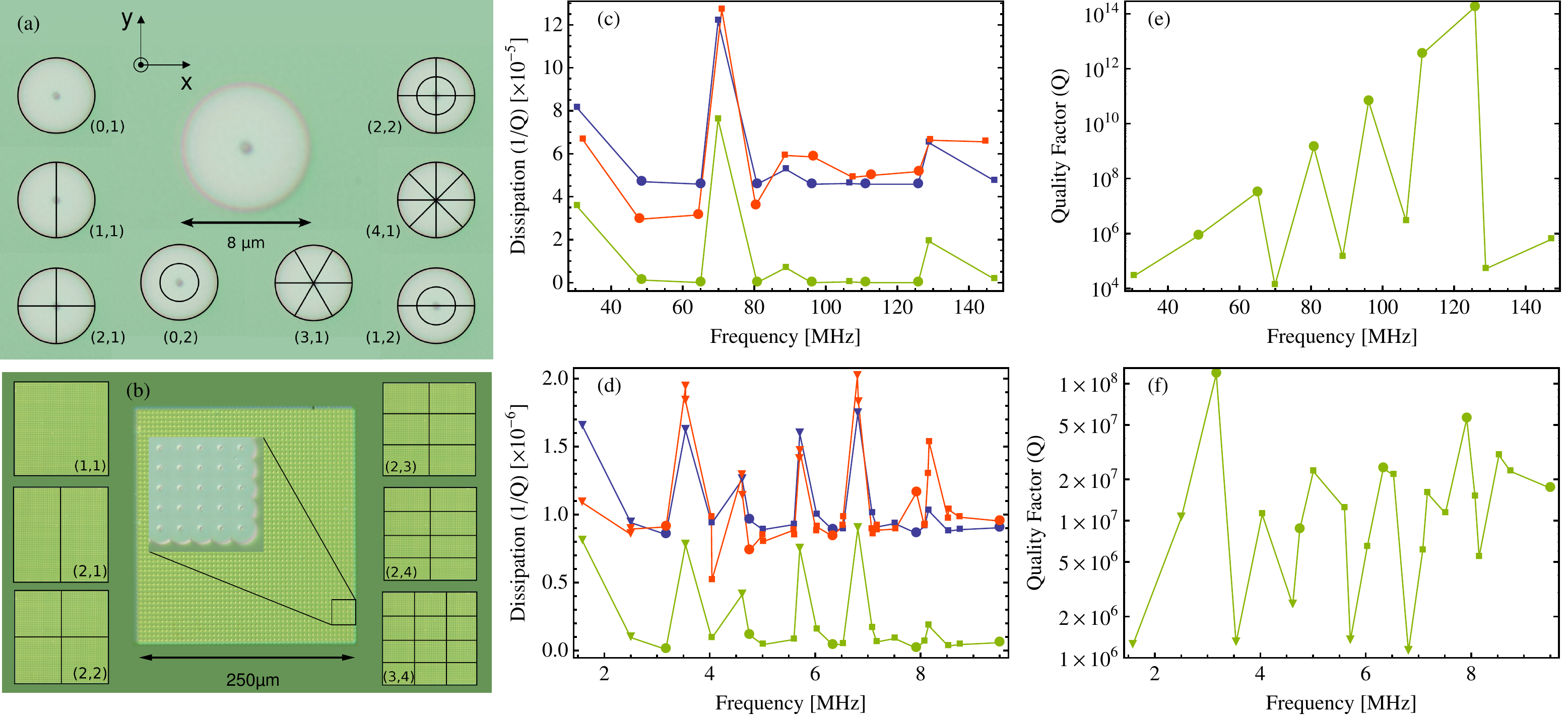}
\caption{(a) Micrograph of a single drum resonator (similar to the one
  used in our analysis) superposed with schematic diagrams of the
  different harmonics $(n,m)$ depicting their nodal lines (the origin
  is set at the center % of mass
  of the membrane). (b) Idem for a square membrane resonator. (c)
  Dissipation $1/Q$ as a function of frequency for the different
  harmonics of a Si$_{3}$N$_{4}$ drum resonator ($D=14.5\,\mu$m,
  $\mathsf{t}=110$nm, and $\sigma=0.90\,$GPa% 0.898285248
  ). (d) Idem for a square membrane resonator
  ($253.2\,\mu$m$\times253.2\,\mu$m$\times0.0125\,\mu$m with
  $\sigma=0.87\,$GPa% 0.867408048
  ). Red plot: Measured values with an error of $10\%$ for $1/Q$ ---
  we ascribe the splitting of degeneracies observed for the square
  membrane [cf.~(d)] to disorder. Blue plot: Least squares fit of our
  model to the measured $1/Q$ using as fit parameters an internal
  dissipation offset ($1/Q_\textrm{int}$) and material properties of
  the substrate ($E_s=148\,$GPa, $\rho_s=3.75$gcm$^{-3}$, and
  $1/Q_\textrm{int}=8.5\times10^{-7}$ for the square membrane, and
  $E_s=323\,$GPa, $1/Q_\textrm{int}=4.6\times10^{-5}$ for the
  drum). Green plot: $1/Q$ without the offset corresponding to the
  predicted clamping loss --- the resulting limits for the $Q$-values
  of the drum and square membrane are shown, respectively, in (e) and
  (f). High-$Q$ harmonics $(n,1)|_{n>0}$ [$(n,n)|_{n>1}$] of the drum
  [square] are marked by circles; low-$Q$ harmonics $(n,1)|_{n>0}$ of
  the square, by triangles. For the drum [square] all harmonics with
  frequencies below $130\,$MHz [$9\,$MHz] are included except
  $(0,3)$, % and
  $(3,2)$ [$(7,1)$%  --- this harmonic, which lies outside the range of
%   validity of Eq.~(\ref{Qsq}), was also measured and distinguished
%   from its accidentally degenerate partner $(5,5)$ by its much lower
%   $Q$
  ].}
\end{figure*}

To derive an adequate model for the clamping losses, we adopt the
phonon tunneling approach introduced in Ref.~\onlinecite{Wilson-Rae08}
and start from the general weak coupling expression for the
dissipation $1/Q$ in terms of the ``overlaps'' between the resonator
mode and the free modes of the substrate (``support'')
\footnote{Equation (\ref{Q}) is also valid for degenerate resonator
  modes $\bar{u}'_{R}$ provided that: (i) the degeneracy (possibly
  split by disorder) is associated to a symmetry also satisfied by the
  support \cite{Wilson-Rae08} or (ii) mode-mixing induced by disorder
  and/or the support is negligible.}:
\begin{align}\label{Q}
  \frac{1}{Q} \,=\, & \frac{\pi}{2 \rho_s \rho_R\omega^3_R}
  \int_q \left| \int_S \rmd \bar{S} \cdot
    \left(\bv{\sigma}^{(0)}_q \cdot \bar{u}'_R - \bv{\sigma}'_R
      \cdot \bar{u}^{(0)}_q \right) \right|^2 \nonumber \\ &
  \times\delta [\omega_R - \omega(q) ]\,.
\end{align}
Here $\bv{\sigma}'_{R}$ and $\bar{u}'_{R}$ are the stress and
displacement fields associated with the normalized resonator mode,
$\bv{\sigma}^{(0)}_{q}$ and $\bar{u}^{(0)}_{q}$ are the analogous
fields for the continuum of support free modes labeled by $q$
[eigenfrequencies $\omega(q)$], and $\rho_s$ and $\rho_R$ are,
respectively, the densities of the substrate and resonator
materials. In our setting the resonator mode should satisfy clamped
boundary conditions at the resonator-support contact area $S$ while
the unperturbed support modes should satisfy free boundary conditions
implying that only the second term in Eq.~(\ref{Q}) contributes. The
substrate is modelled as a half-space that contacts the membrane
resonator at its rim $S$ --- i.e.~the underetched gap between the
suspended structure and the substrate is neglected when determining
the support free modes \footnote{For our structures this gap was
  $\lesssim200$nm$\,\ll D$.}.  We assume the ``high stress'' regime
$\mathsf{t}^2/D^2\ll \sigma/E_R \ll 1$, where $\sigma$ is the tensile
stress in the membrane, $\mathsf{t}$ its thickness, $D$ its large
dimension (diameter or side) and $E_R$ the Young modulus of the
resonator material. This implies that bending effects are negligible
and one can use the classical wave equation adequate for a taut
membrane \cite{Graff}. Thus for the drum's eigenfrequencies we obtain
[cf.~Fig.~1 (a), (b)]: $\omega_{nm}=2\zeta_{nm}c_R/D$ with
$n=0,1,\ldots$ and $m=1,2,\ldots$; while the square's eigenfrequencies
are given by: $\omega_{nm}=\pi\sqrt{n^2+m^2}c_R/D$ with
$n,m=1,2,\ldots$ Here $c_R=\sqrt{\sigma/\rho_R}$ is the phase velocity
in the membrane, and $\zeta_{nm}$ is the $m$th zero of the Bessel
function $J_n(x)$. In this context, the weak coupling approximation
underpinning Eq.~(\ref{Q}) reads $\omega_{nm}\mathsf{t}/c_R\ll1$.

For the single drum we adopt support eigenmodes
$\bar{u}^{(0)}_{q,\theta,l,\gamma}(\bar{r})$ (with
$l\!\!=\!\!0,\pm1,\ldots$) that have axial symmetry with respect
to $z$ [cf.~Fig.~1 (a)]. These are related to the plane wave
eigenmodes $\bar{u}^{(0)}_{\bar{q},\gamma}(\bar{r})$ by:
$\bar{u}^{(0)}_{q,\theta,l,\gamma}(\bar{r})\!=\![(-i)^n/\sqrt{2\pi}]
\int_{-\pi}^{\pi}\rmd\varphi e^{in\varphi}
\bar{u}^{(0)}_{\bar{q},\gamma} (\bar{r})$; where $\gamma\!=\!l$,
$t$, $s$ labels the different types of relevant modes
[i.e.~longitudinal ($l$), transverse SV ($t$), and SAW ($s$)
given that SH waves do not contribute] with velocities of
propagation $c_\gamma$, and we use spherical coordinates for the
incident wavevector $\bar{q}(q,\theta,\varphi)\!
=\!q(\sin\theta\cos\varphi,\sin\theta\sin\varphi, \cos\theta)$
[$\theta\!=\!\pi/2$ for $\gamma\!=\!s$ and $\theta\leq\pi/2$
otherwise].  We note that the stress $\bv{\sigma}'_{R}$
corresponds to the variation with respect to equilibrium and
that the resonant wavevectors in the substrate satisfy
$\omega_{nm}/c_\gamma\mathsf{t}\!\!\ll\!\!1$, so that we can
neglect the variation of
$\bar{u}^{(0)}_{q,\theta,l,\gamma}(\bar{r})$ across the
thickness $\mathsf{t}$. Thus, substitution of the support and
resonator modes into Eq.~(\ref{Q}) (cf.~Appendix \ref{app:dr})
leads to
\begin{equation}\label{Qdr}
  \frac{1}{Q_{nm}}=\frac{4\pi^2\zeta_{nm}\rho_R
    \mathsf{t}}{\rho_s D}\sum_\gamma\eta_\gamma^3
  \tilde{u}_{n,\gamma}(\eta_\gamma\zeta_{nm},\nu_s)\,.   
\end{equation} 
Here we introduce the dimensionless functions
$\tilde{u}_{l,\gamma\neq s}(\tilde
q,\nu_s)\!=\!2\pi\int_0^{\pi/2}\rmd \theta \sin\theta
|u^{(0)}_{\bar q,\gamma;z}(0,\nu_s)|^2J^2_l(\tilde q
\sin\theta)$, $\tilde{u}_{l,s}(\tilde q,\nu_s)\!=\!2\pi
|u^{(0)}_{\bar q,s;z}(0,\nu_s)|^2J^2_l(\tilde q)$ and define
$\eta_\gamma\!\equiv\!
c_R/c_\gamma\!\sim\!\sqrt{\sigma\rho_s/E_s\rho_R}$ --- where the
prefactors of order unity, which depend on $\gamma$, are
functions of the Poisson ratio for the substrate $\nu_s$. We
note that $|u^{(0)}_{\bar q,\gamma;z}(0,\nu_s)|^2$ solely
depends on $\cos\theta$ and $\nu_s$
[cf.~Eqs.~(\ref{u0:l})-(\ref{u0:s})].

In turn, for the square membrane an analogous procedure detailed in
Appendix \ref{app:sq}, leads to:
%\vspace{-0.22cm}
\begin{equation}\label{Qsq}
\frac{1}{Q_{nm}} = \frac{16\pi n^2m^2\rho_R
    \mathsf{t}}{\sqrt{n^2+m^2}\rho_s D} %\sum^\infty_{l=0}
  \sum_{l,\gamma}\eta_\gamma^3
  \tilde{w}^{n,m}_{l,\gamma}(\sqrt{n^2+m^2}\,\eta_\gamma,\nu_s) 
\end{equation}
with $\tilde{w}^{n,m}_{l,s}(\tilde{q},\nu_s) =|u^{(0)}_{\bar
  q,s;z}(0,\nu_s)|^2\tilde{f}_{nml}(\tilde q)$,
$\tilde{w}^{n,m}_{l,\gamma\neq s}(\tilde{q},\nu_s)
=\int_0^{\pi/2}\rmd \theta \sin\theta|u^{(0)}_{\bar
  q,\gamma;z}(0,\nu_s)|^2\tilde{f}_{nml}(\tilde q \sin\theta)$,
where $l\geq0$ and we introduce $\tilde{f}_{nml}(x) =
f_l(\pi x)[Z_{nml}(x)+Z_{mnl}(x)]$ with
%\vspace{-0.22cm}
\begin{align*}
  Z_{nml}(x)\!\equiv &
  \frac{z_{n<}^l(x)}{n^3\left(n^2-x^2\right)^{3/2}}\left\{2n(l+1)
    \sqrt{n^2-x^2} + z_{n<}(x)\right. \\ & \left. +
    \frac{16n^2\left(n^2-x^2\right)z_{n<}(x)}{\left[z_{n<}(x)+z_{m>}(x)\right]
      \left[z_{n<}(x)+z_{m<}(x)\right]}\right\}.
\end{align*}
Here $z_{n\lessgtr}(x)\equiv2n^2-x^2\mp 2n\sqrt{n^2-x^2}$ and the
functions $f_l(x)$ are given by: $f_l(x)\equiv[\delta_{0l}-2(-1)^n
J_{4l}(x)+(-1)^l J_{4l}(\sqrt{2}x)] /(2^{\delta_{0l}}x^{4l})$ for
$n+m$ even, and $f_l(x)=[\delta_{0l}-2\sin^2(l\pi/2)J_{2l}(x)-
\cos(l\pi/2)J_{2l}(\sqrt{2}x)] /(2^{\delta_{0l}}x^{2l})$ for $n+m$
odd. Equation (\ref{Qsq}) is only valid for the case
$\min\{n,m\}>\eta_s\sqrt{n^2+m^2}$ which is satisfied for the
resonances studied here --- note that material properties always imply
$\eta_l<\eta_t<\eta_s$, and $\sigma\ll E_s$ implies $\eta_s\ll1$.

We proceed to compare the predictions of our model
[Eqs.~(\ref{Qdr})-(\ref{Qsq})] with the dissipation measured in
nanomechanical membrane resonators (cf.~Fig.~1). These resonators are
made of ``stoichiometric'' Si$_3$N$_4$ deposited by low pressure
chemical vapor deposition on SiO$_2$ \cite{Southworth09}.  The nitride
has an inherent stress of $1.2\,$GPa, as measured by a wafer bow
technique \cite{Verbridge06}, and a density
$\rho_R=2.7$gcm$^{-3}$. After lithographic patterning to define access
holes, the resonators are suspended by etching the underlying oxide
through these holes, using buffered oxide etch (BOE) for the single
drum and HF for the square membrane, and critical point dried. Thus a
single access hole results in a circular drum geometry, while a square
geometry is defined by a periodic square lattice of such holes
($50\times50$ separated by $5\,\mu$m). Given the small size of the
holes ($\lesssim1\,\mu$m) compared with the typical mode wavelength,
we neglect them in our model. For the square array, the same
consideration applies to the hole separation so that we use a square
membrane model with uniform thickness $\mathsf{t}=12.5\,$nm given by
the average over the array \footnote{The average thickness of the
  membrane is inferred from modelling the isotropic etching process in
  each material.} and side $D=\sqrt{A}=253.2\,\mu$m, where $A$ is the
suspended area. For the single drum (diameter $D=14.5\,\mu$m) the use
of a BOE etch implies that the thickness is uniform and equal to the
nitride thickness ($110$nm).

The mechanical resonances of the structures are characterized under
vacuum and room temperature conditions, using a technique described in
Ref.~\onlinecite{Southworth09}. The resonators are actuated using a
piezo disc that vibrates the chip in the out-of-plane direction and
the motion is detected via a $633$nm continuous wave laser. Figure 1
(c)-(f) compares the measured frequencies and $Q$-values of different
harmonics for the two configurations, single drum and square array,
with the predictions of our model. This comparison takes into account
three issues: (i) the release of the resonator leads to a local
deformation of the wafer that lowers the membrane's tensile stress
with respect to the one in the nitride layer, (ii) in addition to
clamping losses the resonator will be affected by internal
dissipation, and (iii) the parameters for the half-space model of the
substrate must be judiciously chosen.

To deal with (i) we determine the membrane phase velocity $c_R$ from a
suitable linear regression that uses as input the resonator size $D$,
the measured frequencies, and their mode indices which can be
identified from the frequency ratios between the harmonics and the
fundamental mode. We find an excellent correlation that yields
$c_R=576.8\,$ms$^{-1}$ ($566.8\,$ms$^{-1}$) for the drum (square).

Our model is in excellent agreement with the observed trends providing
the internal dissipation channel (ii) is frequency independent, and
can be just added as a fit parameter ($1/Q_\textrm{int}$) to the
calculated dissipation. To elucidate (iii) one needs to compare the
wavelengths of the resonant ``support'' modes with the thickness of
the Si wafer ($0.5$mm). For the square the resonant frequencies are in
the MHz range resulting in wavelengths in Si ($4$-$8$mm) much larger
than the wafer's thickness so that these modes are dominated by the
properties of the underlying piezo and positioning system. Thus, we
adopt $\nu_s=1/3$ and leave the density $\rho_s$ and Young modulus
$E_s$ as fit parameters. On the other hand for the drum the resonances
studied lie in the $100\,$MHz range so that the elastic wave radiation
is determined mostly by the anisotropic properties of crystalline
Si. For this case we adopt $\rho_s=2.33\,$gcm$^{-3}$ and $\nu_s=0.28$,
but leave $E_s$ as a fit parameter given the isotropic nature of our
model.

In both cases, drum and square geometry, we find a class of modes that
consistently exhibit lower dissipation $1/Q$ when compared to nearby
modes [cf.~Fig.~1 (c) and (d)].  Their measured $Q$ remains
approximately constant as the harmonic index is increased, leading to
a growth in their $fQ$ product that for the square reaches a maximum
of $1.0\times10^{13}$Hz for the $(6,6)$ harmonic. These ``special''
classes of harmonics for the drum and square are, respectively,
$(n,1)|_{n>0}$ and $(n,n)|_{n>1}$ and correspond to the presence of
nodal lines that intersect the periphery at evenly spaced points
[cf.~Fig.~1 (a) and (b)]. In contrast, for the square geometry the
modes $(n,1)$, $(1,n)$, where two of the sides do not intersect any
nodal lines, tend to exhibit smaller $Q$ for comparable frequencies
with $fQ\sim10^{12}$Hz. An intuitive heuristic understanding of these
trends emerges from realizing that for low harmonics, with membrane
wavevectors $\sim\pi/D$, the typical resonant wavelengths in the
substrate are much larger than $D$. Thus, for the special modes the
clamping loss is suppressed [cf.~Fig.~1 (e) and (f)] due to
destructive interference between the waves radiated by the different
equivalent segments of the periphery, defined by the nodal lines,
which have alternating $\pi$-phases. Concomitantly, unlike the
fundamental mode, these special modes are associated to stress sources
with vanishing total force. 

A quantitative grasp of these striking features can be gained by
exploiting the smallness of the $\eta_\gamma$ underpinning the
aforementioned wavelength separation. For the drum, relevant harmonics
satisfy the condition $\eta_\gamma\zeta_{nm}\ll\sqrt{n+1}$ which
allows us to Taylor expand the Bessel functions in the
$\tilde{u}_{l,\gamma}$ yielding an approximation for Eq.~(\ref{Qdr})
that implies the following \footnote{The approximate scaling for
  $Q_{n1}/Q_{01}$ differs from the results of applying Eq.~(\ref{Qdr})
  by at most a factor of $2$ for $0<n<15$ in the relevant regime
  $\eta_t<0.3$ and $\nu=1/3$.}: %\vspace{-0.3cm}
\begin{align}\label{scaling:dr}
  Q_{01} \,\approx\,\, & \left.\frac{\rho_sc_t^3}{2\pi^2
      \sigma_Rc_R^2\omega_{01}\tilde{u}_0 (\nu_s)}
  \right|_{\nu_s=1/3}\!\!\!\!\!\!\!=0.029
  \sqrt{\frac{\rho_R}{\rho_s}\left(\frac{E_s}{\sigma}\right)^3}
  \frac{D}{\mathsf{t}} \nonumber \\ %\label{scaling:drn}
  \frac{Q_{n1}}{Q_{01}} \,\sim\,\, & n^{\frac{2n+1}{16}}
  \!\left(0.517\frac{c_s}{c_R}\right)^{2n}, \ 
  \frac{Q_{nm}}{Q_{n1}}\approx
  \left(\frac{\zeta_{n1}}{\zeta_{nm}}\right)^{2n+1}
\end{align}
where $\sigma_R=\rho_R\mathsf{t}$ is the surface mass density of the
membrane and $\tilde{u}_0 (\nu_s)\equiv\sum_\gamma(c_t/c_\gamma)^3
\tilde{u}_{0,\gamma}(0,\nu_s)$. Thus, the clamping-loss limited
$Q$-values of modes $(n,1)$ effectively grow exponentially --- as the
super-exponential factor plays a negligible role for relevant $n$, in
sharp contrast to series of modes for which $m$ is increased while $n$
is kept constant. These exhibit a decrease of $Q_\textrm{clamp}$ for
increasing frequency \footnote{For such modes,
  $\eta_\gamma\zeta_{nm}\ll\sqrt{n+1}$ eventually fails and for
  $\eta_\gamma\zeta_{nm}\gg |n^2-1/4|$ one finds that $1/Q_{nm}$
  exhibits oscillatory behavior dominated by the SAW
  contribution.}. On the other hand, for the square geometry analogous
considerations imply for $m\sim n\gg\zeta_{01}/2\pi\eta_\gamma$ a rise
in $Q_\textrm{clamp}$ that is merely linear, with the damping rate
tending to a constant value, as the harmonic indices are increased
with their ratio $m/n$ fixed.  In turn, for our setting given the
magnitude of $1/Q_\textrm{int}$ all the high-$Q$ modes present roughly
constant $Q$-values.

A comparison between the predictions [cf.~Eqs.~(\ref{Qdr}) and
(\ref{Qsq})] for the two geometries (with appropriate dimensions) also
reveals that for ``special'' harmonics [$(n,1)|_{n>0}$ and
$(n,n)|_{n>1}$] with the same frequency and number of nodal lines the
circular geometry always yields a higher $Q$.  Finally, one should
note that the scalings, embodied in Eq.~(\ref{scaling:dr}), for the
$Q$-values in terms of $\rho_R/\rho_s$, $E_s/\sigma$, and
$D/\mathsf{t}$ are completely general and independent of the shape of
the boundary. These, directly imply that the $fQ_\textrm{clamp}$
product of a given harmonic is independent of $D$. Furthermore,
typical parameters yield for the fundamental mode
$fQ_\textrm{clamp}\sim10^{12}$Hz, which is comparable to experimental
values (cf.~Fig.~1 (c) and (d), and
Refs.~\onlinecite{Thompson08}, \onlinecite{Wilson09}).

In conclusion, we find that the dissipation of different harmonics of
a given membrane resonator exhibit a striking non-monotonic behavior
which can be understood in terms of how the mode-shapes of different
harmonics influence the clamping loss. We find classes of modes for
which the measured $Q$ remains approximately constant and
substantially larger than for other modes with comparable frequency,
and explain this phenomenon in terms of destructive interference
between the radiated waves leading to a strong suppression of the
clamping loss. Notably, our analysis implies that for modes $(n,1)$ of
a circular geometry, the damping rate due to elastic-wave radiation
vanishes exponentially in $n$ rendering them ``asymptotically mute''.
Thus, for typical parameters, these azimuthal harmonics can be
regarded as effectively clamping-loss free for moderate $n$
(e.g.~$fQ_\textrm{clamp}\gtrsim10^{17}$Hz for $n\geq5$ and thickness
$\mathsf{t}<100\,$nm). % (e.g.~$fQ_\textrm{clamp}\gtrsim10^{18}$Hz for
% $n\geq6$ and thickness $\mathsf{t}<250\,$nm).
Our results are relevant to state-of-the-art dispersive optomechanical
setups \cite{Thompson08,Wilson09} and the model is also applicable to
graphene nanodrums under tension \cite{Chen09}.  Finally, we highlight
that the interference effects we have unveiled will also be
operational for the flexural modes of rigid plates.

IWR acknowledges financial support via the Nanosystems Initiative
Munich. Work at Cornell was supported under NSF ECCS 1001742.

\appendix

\section{Circular geometry}\label{app:dr}

In the case of the circular geometry, the axially symmetric support
eigenmodes used are related to the plane wave eigenmodes
$\bar{u}^{(0)}_{\bar{q},\gamma}(\bar{r})$ by:
\begin{equation}\label{cylin}
  \bar{u}^{(0)}_{q,\theta,l,\gamma}(\bar{r})=\frac{(-i)^n}{\sqrt{2\pi}}
  \int_{-\pi}^{\pi}\rmd\varphi e^{in\varphi}
  \bar{u}^{(0)}_{\bar{q}(q,\theta,\varphi),\gamma} (\bar{r})\,. 
\end{equation}
Here $\gamma=l$, $t$, $s$ labels the different types of relevant
modes [i.e.~longitudinal ($l$), transverse SV ($t$), and SAW
($s$) given that SH waves do not contribute] and
$\bar{q}(q,\theta,\varphi)
=q(\sin\theta\cos\varphi,\sin\theta\sin\varphi, \cos\theta)$ is
the incident wavevector [$\theta=\pi/2$ for $\gamma=s$ and
$\theta\leq\pi/2$ otherwise]. As the resonator material is
prestressed, the stress $\bv{\sigma}'_{R}$ in Eq.~(1) is not the
total stress but corresponds instead to the variation with
respect to equilibrium. This, together with the validity of
membrane theory, \cite{Graff} implies that to linear order in
the displacement field we have
\begin{equation}\label{stress1}
\rmd\bar{S}\cdot\bv{\sigma}'_{R} \,\parallel\,\hat z
\end{equation}
and
\begin{equation}\label{stress2}
\int^{\mathsf{t}/2}_{-\mathsf{t}/2}\hat{z}\cdot\bv{\sigma}'_{R}
\cdot\hat{r}\rmd z = \sigma\sqrt{\mathsf{t}}
\frac{\partial\phi_R}{\partial r}\,,
\end{equation}
which we will use to simplify Eq.~(1). Here we introduce the
normalized resonator eigenmodes $\phi_R\to\phi_{nm}$, that
satisfy the 2D classical wave equation with $\phi_{nm}=0$ at the
periphery of the membrane, and adopt cylindrical coordinates
$\bar r\!=\!(r\cos\phi,r\sin\phi,z)$. These eigenmodes read
\begin{align}\label{reson1}
  \phi_{0m}(r,\phi)& =\sqrt{\frac{1}{\pi}}R_{0m}(r)\,,\nonumber\\
  \phi_{nm}(r,\phi)& =\sqrt{\frac{2}{\pi}}R_{nm}(r)\cos n\phi\,, \nonumber\\
  \tilde{\phi}_{nm}(r,\phi) & = \sqrt{\frac{2}{\pi}}R_{nm}(r)\sin
  n\phi
\end{align}
with
\begin{equation}\label{reson2}
  R_{nm}(r) = \frac{2J_n(2\zeta_{nm}r/D)}{D J_{n+1}(\zeta_{nm})}\,,
\end{equation}
and have eigenfrequencies $\omega_R\to\omega_{nm}$ given by
\begin{equation}\label{freq:dr}
\omega_{nm}=2\zeta_{nm}\frac{c_R}{D}\,.
\vspace{0.2cm}
\end{equation}
Here $n=0,1,\ldots$, $m=1,2,\ldots$, and the harmonics with $n>0$
are doubly degenerate.

As the resonant wavevectors in the substrate satisfy
$\omega_{nm}/c_\gamma\mathsf{t}\ll1$ we can neglect the
variation of $\bar{u}^{(0)}_{q,\theta,l,\gamma}(\bar{r})$ across
the thickness $\mathsf{t}$ (i.e.~the $z$-dependence at
$S$). This approximation and Eq.~(\ref{stress1}) imply that the
support modes only enter into Eq.~(1) through
$u^{(0)}_{q,\theta,l,\gamma;z}(\bar{r})|_{z=0}$. To determine
the latter we exploit that reflection at the free surface
preserves the tangential component of the wavevector implying
\begin{equation}\label{u:z}
  u^{(0)}_{\bar{q}(q,\theta,\varphi),\gamma;z}(\bar{r})|_{z=0}
  =u^{(0)}_{\bar{q}(q,\theta,\varphi),\gamma;z}(0)
  e^{irq\sin\theta\cos(\phi-\varphi)}\,.
\end{equation}
Then, from Eqs.~(\ref{cylin}) and (\ref{u:z}), using the Bessel
integral $J_n(x)=\frac{1}{2\pi}
\int_{-\pi}^{\pi}e^{-i(n\phi-x\sin\phi)}\rmd\phi$, we obtain
\begin{equation}\label{cylin:z}
u^{(0)}_{q,\theta,l,\gamma;z}(\bar{r})|_{z=0}= \sqrt{2\pi}
u^{(0)}_{\bar{q},\gamma;z}\!(0) J_n(rq\sin\theta) e^{i n\phi}\,.
\end{equation}  
We now deploy Eq.~(\ref{stress1}) and substitute
Eqs.~(\ref{stress2})-(\ref{reson2}) and (\ref{cylin:z}) into
Eq.~(1). Subsequently, we use that here
\begin{equation}
\int_q\to\sum_{l,\gamma} \int_0^{\infty}\!\rmd q q^{d_\gamma-1}\left[\left(1-
\delta_{\gamma s}\right)\int_0^{\pi/2}\!\rmd\theta\sin\theta +
\delta_{\gamma s}\right]\nonumber
\end{equation}
where $d_\gamma$ is the dimensionality (i.e.~$d_\gamma=2$ for
$\gamma=s$ and $d_\gamma=3$ for $\gamma\neq s$), perform the
substitution $\omega\!=\!c_\gamma q$ (for each $\gamma$),
evaluate $\frac{\partial\phi_R}{\partial r}$ using
$J'_n(\zeta_{nm})\!=\!-J_{n+1}(\zeta_{nm})$, and express
$\sigma$ in terms of $c_R$ and $\rho_R$. Finally, integration
over $\omega$ and $\phi$, summation over $l$ (all terms vanish
except $l\!=\!\pm n$), and substitution of Eq.~(\ref{freq:dr})
yields Eq.~(2).

The $z$-displacements at the origin
$u^{(0)}_{\bar{q}\gamma;z}(0)$ only depend on $\cos\theta$ and
$\nu_s$, and their absolute values squared are given by the
following \cite{Wilson-Rae08}
\begin{widetext}
\begin{align}
  \left.|u^{(0)}_{\bar{q}(q,\theta,\varphi),l;z}(0,\nu_s)|^2\right|_{\cos\theta=v}=\,
  & \left. \frac{(1 - 2 \alpha + 2 \alpha v^2)^2
      v^2}{2\pi^3\left[ 4 \alpha^{3/2} \sqrt{1-\alpha+\alpha
          v^2} (1-v^2) v + (1 - 2 \alpha + 2 \alpha v^2)^2
      \right]^2}\right|_{\alpha=\frac{1-2\nu_s}{2(1-\nu_s)}}\,, \label{u0:l}\\
  \left.|u^{(0)}_{\bar{q}(q,\theta,\varphi),t;z}(0,\nu_s)|^2\right|_{\cos\theta=v}
  =\, & \left.\frac{2(1-\alpha-v^2) (1-v^2) v^2}{\pi^3\left[16
        (1-\alpha-v^2) (1-v^2)^2 v^2 + (2
        v^2-1)^4\right]}\right|_{\alpha=\frac{1-2\nu_s}{2(1-\nu_s)}}\qquad
  \mathrm{for}\ 0<v<\sqrt{1-\alpha}\,,
  \label{u0:t1}\\
  \left.|u^{(0)}_{\bar{q}(q,\theta,\varphi),t;z}(0,\nu_s)|^2\right|_{\cos\theta=v}
  =\, & \left.\frac{2(\alpha-1+v^2) (1-v^2) v^2}{\pi^3\left[4
        \sqrt{\alpha-1+v^2} (1-v^2) v + (2 v^2-1)^2
      \right]^2}\right|_{\alpha=\frac{1-2\nu_s}{2(1-\nu_s)}}\ \ \qquad
  \mathrm{for}\ \sqrt{1-\alpha}<v<1\,,
  \label{u0:t2}\\
 \left.|u^{(0)}_{\bar{q}(q,\theta,\varphi),s;z}(0,\nu_s)|^2\right|_{\cos\theta=v} = &
  \left.\frac{C^2(\alpha)}{2\pi^2}\left[\sqrt{1-\alpha
        \xi^2\!(\alpha)} -
      \frac{1-\xi^2\!(\alpha)/2}{\sqrt{1-\xi^2\!(\alpha)}}
    \right]^{2}\right|_{\alpha=\frac{1-2\nu_s}{2(1-\nu_s)}}\,.\label{u0:s}
\end{align}
\end{widetext}

\section{Square geometry}\label{app:sq}

For the square geometry we exploit that the whole structure
still presents reflection symmetries with respect to the $x-z$
and $y-z$ planes. These are associated, respectively, with the
operators $\hat R_x$ and $\hat R_y$ acting on the space of
solutions of the elastic wave equations. Thus we can use normal
modes of the decoupled support (elastic half-space) and of the
resonator (i.e.~square membrane) that are eigenvectors of $\hat
R_x$ and $\hat R_y$. In the case of the support, for a given
plane wave mode $|u\rangle\doteq
u^{(0)}_{\bar{q}\gamma;z}(\bar{r})$ of the elastic half-space
one can generate modes $|u_{++} \rangle$, $|u_{+-}\rangle$,
$|u_{-+} \rangle$, and $|u_{--} \rangle$ with the desired
reflection properties by the following symmetrization procedure
\begin{widetext}
\begin{equation}\label{eq:sym}
| u_{\mu\nu} \rangle\! =\! \hat S_{\mu\nu} | u \rangle
 \!\equiv\!\frac{1}{2}\!\left(\! | u \rangle + \mu \hat R_x | u \rangle
 + \nu \hat R_y | u \rangle+ \mu \nu\hat R_x \hat R_y | u \rangle
\! \right)
\end{equation}
with $\mu,\nu = \pm$, which enforces
\begin{align}\label{refsym}
&\hat R_x | u_{\mu\nu} \rangle = \mu | u_{\mu\nu} \rangle \,,
\nonumber\\ &\hat R_y | u_{\mu\nu} \rangle = \nu | u_{\mu\nu} \rangle
\,.
\end{align}
Naturally a complete basis is obtained by taking $\bar{q}$ in
the first quadrant, i.e.~$0\leq\varphi\leq\pi/2$.  As for the
drum: (i) the smallness of $\mathsf{t}$ compared with the
resonant wavelengths in the substrate implies that we can
neglect in Eq.~(1) the $z$-dependence of the symmetrized support
modes at $S$, and (ii) only their $z$-components are
relevant. These $z$-components read
\begin{align}
  u^{(0)}_{\bar{q}\gamma,++;z}(x,y,0) & =  2
  u^{(0)}_{\bar{q}\gamma;z}(0)\cos(qx\sin\theta\cos\varphi)
  \cos(qy\sin\theta\sin\varphi)\,,
 \label{support:sym}% \nonumber
\\
  u^{(0)}_{\bar{q}\gamma,--;z}(x,y,0) & = 2
  u^{(0)}_{\bar{q}\gamma;z}(0)\sin(qx\sin\theta\cos\varphi)
  \sin(qy\sin\theta\sin\varphi)\,,
 \label{support:antisym}% \nonumber
\\
  u^{(0)}_{\bar{q}\gamma,+-;z}(x,y,0) & = 2
  u^{(0)}_{\bar{q}\gamma;z}(0)\cos(qx\sin\theta\cos\varphi)
  \sin(qy\sin\theta\sin\varphi)\,,
 \label{support:mixed1}% \nonumber
\\
  u^{(0)}_{\bar{q}\gamma,-+;z}(x,y,0) & = 2
  u^{(0)}_{\bar{q}\gamma;z}(0)\sin(qx\sin\theta\cos\varphi)
  \cos(qy\sin\theta\sin\varphi)\,.\label{support:mixed2}
\end{align}
In turn, for the resonator modes we obtain
\begin{align}
  \phi_{n,m}(x,y) & =\frac{2}{D}\cos\frac{n\pi}{D}x\cos\frac{m\pi}{D}y
 % \ \qquad 
&& \mathrm{for}\ n,m\ \mathrm{odd}\ \to\  \mathrm{symmetric\ case}\,,
\label{resonator:sym}%  \nonumber
\\
  \phi_{n,m}(x,y) & =\frac{2}{D}\sin\frac{n\pi}{D}x\sin\frac{m\pi}{D}y
%  \ \qquad 
&& \mathrm{for}\ n,m\ \mathrm{even}\ \to\  \mathrm{antisymmetric\ case}\,,
\label{resonator:antisym}%  \nonumber
\\
  \phi_{n,m}(x,y) & =\frac{2}{D}\cos\frac{n\pi}{D}x\sin\frac{m\pi}{D}y
%  \ \qquad 
&& \mathrm{for}\ n\ \mathrm{odd\ and}\ m\ \mathrm{even}\ \to\ 
\mathrm{mixed\ symmetry\ case}\,, 
\label{resonator:mixed1}%  \nonumber
\\
  \phi_{n,m}(x,y) & =\frac{2}{D}\sin\frac{n\pi}{D}x\cos\frac{m\pi}{D}y
%  \ \qquad 
&& \mathrm{for}\ n\ \mathrm{even\ and}\ m\ \mathrm{odd}\ \to\  \mathrm{mixed\
  symmetry\ case}\,, 
\label{resonator:mixed2}
\end{align}
%\end{widetext}
with eigenfrequencies 
\begin{equation}\label{freq:sq}
  \omega_{nm}=\pi\sqrt{n^2+m^2}\frac{c_R}{D}\qquad\qquad\ n,m=1,2,\ldots
\end{equation}
Naturally only modes with the same reflection symmetries are
coupled via Eq.~(1). We first treat the symmetric modes and then
briefly outline the straightforward extension to the other
cases.

\subsection{Symmetric modes}\label{subsec:symm}

We make use of
\begin{equation}\label{trigonometric}
  \int_{-D/2}^{D/2}\rmd x\cos\frac{n\pi}{D}x\cos q_x x =
  \frac{\sin\tfrac{n\pi+q_xD}{2}}{\frac{n\pi}{D}+q_x}+
  \frac{\sin\tfrac{n\pi-q_xD}{2}}{\frac{n\pi}{D}-q_x}\,, 
\end{equation}
Eq.~(\ref{stress2}), and the symmetries of the square, to obtain
from Eqs.~(\ref{support:sym}) and (\ref{resonator:sym}) the
following
%\begin{widetext}
\begin{align}\label{intphi1}
  \int_0^{\pi/2}\rmd\varphi\left|\int_S\rmd\bar
    S\cdot\bv{\sigma}_R\cdot\bar{u}^{(0)}_{\bar{q}\gamma,++}\right|^2=\,\, &
  256\pi^4n^2m^2\frac{\sigma^2\mathsf{t}}{D^2} \left|
    u^{(0)}_{\bar{q}\gamma;z}(0)\right|^2
  \int_0^{\pi/2}\rmd\varphi
  \cos^2\left(\tfrac{qD}{2}\sin\theta\cos\varphi\right)
  \cos^2\left(\tfrac{qD}{2}\sin\theta\sin\varphi\right)\nonumber\\
  &\times \left[\frac{1}{(n\pi)^2-(qD\sin\theta\cos\varphi)^2}+
    \frac{1}{(m\pi)^2-(qD\sin\theta\sin\varphi)^2}\right]^2\,.
\end{align}
Subsequently we substitute into Eq.~(\ref{intphi1}) the
decomposition
\begin{align}\label{intphi2}
  4\cos^2\left(\tfrac{qD}{2}\sin\theta\cos\varphi\right)
  \cos^2\left(\tfrac{qD}{2}\sin\theta\sin\varphi\right) = &\, 1+
  \cos(qD\sin\theta\cos\varphi) +
  \cos[qD\sin\theta\cos(\varphi-\tfrac{\pi}{2})]\nonumber\\ &
  +\tfrac{1}{2}\cos[\sqrt{2}qD\sin\theta\cos(\varphi-\tfrac{\pi}{4})]
  +\tfrac{1}{2}\cos[\sqrt{2}qD\sin\theta\cos(\varphi+\tfrac{\pi}{4})]
  \nonumber\\ = &\, 1+2J_0(qD\sin\theta)+4\sum_{l=1}^\infty
  J_{4l}(qD\sin\theta)\cos4l\varphi \nonumber\\  & +
  J_0(\sqrt{2}qD\sin\theta)+2\sum_{l=1}^\infty(-1)^l
  J_{4l}(\sqrt{2}qD\sin\theta)\cos4l\varphi\,,
\end{align}
which follows from the Jacobi-Anger expansion $e^{i
  x\cos\phi}=\sum_{n=-\infty}^\infty i^nJ_n(x)e^{i n\phi}$ and
the relation $J_{-n}(x)=(-1)^nJ_{n}(x)$. Then, using the
invariance of the integrand under $\varphi\to\pi-\varphi$, which
implies $\int_0^{\pi/2}\to\tfrac{1}{2}\int_0^{\pi}$, and
performing the substitution $\varphi'=2\varphi$ we obtain
\begin{align}\label{intphi3}
  \int_0^{\pi/2}\rmd\varphi\left|\int_S\rmd\bar
    S\cdot\bv{\sigma}_R\cdot\bar{u}^{(0)}_{\bar{q}\gamma,++}\right|^2=\,\,
  &
  64n^2m^2\frac{\sigma^2\mathsf{t}}{D^2}\left|u^{(0)}_{\bar{q}\gamma;z}(0)
  \right|^2\left(\frac{\pi}{qD\sin\theta}\right)^4
  \Re\left\{\int_0^{2\pi}\rmd\varphi
    \sum_{l=0}^\infty a_l(q D\sin\theta)e^{i 2l\varphi}\right.\nonumber\\
  &\left.\times \left[\frac{1}{2\left(\frac{n\pi}{q
            D\sin\theta}\right)^2-1-\cos\varphi}+
      \frac{1}{2\left(\frac{m\pi}{q
            D\sin\theta}\right)^2-1+\cos\varphi}\right]^2\right\}%\,.
\end{align}
with
\begin{equation}\label{a}
  a_0(x)=1+2J_{0}(x)+J_{0}(\sqrt{2}x)\,, \qquad\quad
  a_l(x)=2\left[2J_{4l}(x)+(-1)^lJ_{4l}(\sqrt{2}x)\right]\quad \
  \mathrm{for}\ l>0\,. 
\end{equation}
%\end{widetext}
The angular integral can be converted into a contour integral
over the unit circle in the complex plane using
$z=e^{i\varphi}$. We focus on the case
$\min\{n,m\}/\sqrt{n^2+m^2}>\eta_s$ in which for all $\gamma$;
the series in Eq.~(\ref{intphi3}) converges uniformly, so that
the integral and the sum commute, and the denominator of the
first (second) term in the last factor has one real root
$\tilde{z}_{n<}$ $(-\tilde{z}_{m<})$ inside the unit circle and
another one $\tilde{z}_{n>}$ $(-\tilde{z}_{m>})$ outside of it
given by \footnote{In the opposite case
  $\min\{n,m\}/\sqrt{n^2+m^2}\leq\eta_s$ there are one or more
  types of modes $\gamma$ for which at least one of the
  denominators in the last factor in Eq.~(\ref{intphi3}) has a
  pair of complex conjugate roots on the unit circle, so that
  this factor diverges for certain angles $\phi$ and the
  convergence of the series is no longer uniform, rendering the
  decomposition used invalid.}
\begin{equation}\label{pole}
  \tilde{z}_{n\lessgtr}= 2\left(\frac{n\pi}{qD\sin\theta}\right)^2-1\mp
  \frac{2n\pi}{qD\sin\theta}\sqrt{\left(\frac{n\pi}{qD\sin\theta}\right)^2-1}
  \,. 
\end{equation}
We now evaluate the resulting integral for each term in $l$ by
residues to obtain
%\begin{widetext}
\begin{align}\label{intphi4}
  \int_0^{\pi/2}\!\!\rmd\varphi\left|\int_S\rmd\bar
    S\cdot\bv{\sigma}_R\cdot\bar{u}^{(0)}_{\bar{q}\gamma,++}\right|^2\!=\,\,
  & 32\pi
  n^2m^2\frac{\sigma^2\mathsf{t}}{D^2}\left|u^{(0)}_{\bar{q}\gamma;z}(0)
  \right|^2\!\left(\frac{2\pi}{qD\sin\theta}\right)^4
  \sum_{l=0}^\infty a_l(q D\sin\theta)\!\left[\frac{(2l+1)
      \tilde{z}_{n<}^{2l}}{\left(\tilde{z}_{n<}-\tilde{z}_{n>}\right)^2}
    -\frac{2\tilde{z}_{n<}^{2l+1}}{\left(\tilde{z}_{n<}-
        \tilde{z}_{n>}\right)^3}\right.
  \nonumber\\
  & \left. + \frac{(2l+1)
      \tilde{z}_{m<}^{2l}}{\left(\tilde{z}_{m<}
        -\tilde{z}_{m>}\right)^2} -\frac{2
      \tilde{z}_{m<}^{2l+1}}{\left(\tilde{z}_{m<}-\tilde{z}_{m>}\right)^3}
    -\frac{2
      \tilde{z}_{n<}^{2l+1}}{\left(\tilde{z}_{n<}-\tilde{z}_{n>}\right)
      \left(\tilde{z}_{n<}+\tilde{z}_{m>}\right)
      \left(\tilde{z}_{n<}+\tilde{z}_{m<}\right)}\right.
  \nonumber\\
  & \left.  -\frac{2
      \tilde{z}_{m<}^{2l+1}}{\left(\tilde{z}_{m<}-\tilde{z}_{m>}\right)
      \left(\tilde{z}_{m<}+\tilde{z}_{n>}\right)
      \left(\tilde{z}_{m<}+\tilde{z}_{n<}\right)}\right]%\,.
\end{align}
\end{widetext}
where we have omitted the dependence on $q D\sin\theta$ of the
last factor. Subsequently, we regroup terms in the above using
Eq.~(\ref{pole}) and substitute Eqs.~(\ref{intphi4}) and
(\ref{a}) into Eq.~(1). Then, we use that here $\int_q$
corresponds to
\begin{equation}
\sum_\gamma\int_0^{\pi/2}\!\rmd\varphi\int_0^{\infty}\!\rmd q
q^{d_\gamma-1}\left[\left(1- \delta_{\gamma
  s}\right)\!\int_0^{\pi/2}\!\rmd\theta\sin\theta + \delta_{\gamma
  s}\right]\,, \nonumber
\end{equation}
perform the substitution $\omega\!=\!c_\gamma q$ (for
each $\gamma$), and integrate over $\omega$. Finally, we
eliminate $\sigma$ in favor of $c_R$ and $\rho_R$, introduce
the functions $f_l(x)\equiv a_l(x)/(2x^{2l})$ and
\begin{equation}\label{z}
  z_{n\lessgtr}(x)\equiv\left.\left(\frac{qD}{\pi}\sin\theta\right)^2
    \tilde{z}_{n\lessgtr}\right|_{\frac{qD}{\pi}\sin\theta=x}\,, 
\end{equation}
and substitute Eq.~(\ref{freq:sq}), to obtain Eq.~(3)
specialized for $n$, $m$ odd.

\subsection{General case}

For the other cases, antisymmetric and mixed-symmetry modes,
analogous steps as for the symmetric modes
(cf.~\ref{trigonometric}), allow us to obtain from
Eqs.~(\ref{stress2}), (\ref{support:antisym}),
(\ref{support:mixed1}), (\ref{resonator:antisym}), and
(\ref{resonator:mixed1}) expressions that differ from
Eq.~(\ref{intphi1}) only in the first two factors of the
R.H.S. integrand which are modified via
\begin{equation}\label{factor:antisym}
 \cos^2(q'_x)\cos^2(q'_y)\to\sin^2(q'_x)\sin^2(q'_y)
\end{equation}
for $n$, $m$ even (antisymmetric modes) and
\begin{equation}\label{factor:mixed1}
 \cos^2(q'_x)\cos^2(q'_y)\to\cos^2(q'_x)\sin^2(q'_y)
\end{equation}
for $n$ odd and $m$ even (mixed symmetry modes), where
$q'_x=\tfrac{qD}{2}\sin\theta\cos\varphi$ and
$q'_y=\tfrac{qD}{2}\sin\theta\sin\varphi$. The results for the
other mixed symmetry case [Eqs.~(\ref{support:mixed2}),
(\ref{resonator:mixed2})] can be obtained by interchanging the
harmonic indices. Subsequently, we express these factors
[Eqs.~(\ref{factor:antisym}), (\ref{factor:mixed1})] as
trigonometric series in $\varphi$ using the Jacobi-Anger
expansion and obtain again Eq.~(\ref{intphi3}), with 
\begin{align}
  a_0(x)& = 1-2J_{0}(x)+J_{0}(\sqrt{2}x)\,, \nonumber\\
  a_l(x)& =-2\left[2J_{4l}(x)-(-1)^lJ_{4l}(\sqrt{2}x)\right]\qquad
  \mathrm{for}\ l>0\,,
\end{align}
in the antisymmetric case, and 
\begin{align}
  a_0(x)& = 1-J_{0}(\sqrt{2}x)\,,\nonumber\\
  a_l(x)& = -4 J_{2l}(x)
  &&\mathrm{for}\ l\ \mathrm{odd}\,,\nonumber\\
  a_l(x)& =-2(-1)^{l/2}J_{2l}(\sqrt{2}x)&&
  \mathrm{for}\ l>0\,,\ \mathrm{even}
\end{align}
in the case of mixed symmetry. Henceforth, the same steps
followed in Sec.~\ref{subsec:symm} [cf.~Eqs.~(\ref{intphi3}),
(\ref{pole})-(\ref{z})] lead to Eq.~(3).

The series in $l$, corresponding to the last dimensionless
factor in Eq.~(3), can be evaluated numerically using
Eqs.~(\ref{u0:l})-(\ref{u0:s}). Without loss of generality, we
assume $n<m$ and find that, for ratios $m/n$ not too large, the
condition $\eta_\gamma\sqrt{1+m^2/n^2}\ll1$ implies that
prevalently
$|\tilde{w}^{n,m}_{l+1,\gamma}/\tilde{w}^{n,m}_{l,\gamma}|\sim\eta_\gamma^2(1+m^2/n^2)/4$
ensuring rapid convergence.

\bibliographystyle{apsrev}
%\bibliography{highQ_interference_etal}
%\bibliography{}

%\vspace{-0.7cm}

\end{document}